\RequirePackage{fix-cm}
\documentclass[smallextended]{svjour3}       
\smartqed  
\usepackage{graphicx}
\usepackage{color}
\begin{document}

\title{Physics Beyond the Multiverse
}
\subtitle{Naturalness and the Quest for a Fundamental Theory}


\author{Heinrich P\"as 
}


\institute{
Fakult\"at f\"ur Physik, Technische Universit\"at Dortmund, 44221 Dortmund, Germany\\
              \email{heinrich.paes@tu-dortmund.de}           
}

\date{Received: date / Accepted: date}

\maketitle

\begin{abstract}
Finetuning and Naturalness are extra-empirical theory assessments that reflect our expectation how scientific theories should provide an intuitive understanding about the foundations underlying the observed phenomena. Recently, the absence of new physics at the LHC and the theoretical evidence
for a multiverse of alternative physical realities, predicted by our best fundamental theories, have casted doubts about the validity of these concepts. In this essay we argue that the
discussion about Finetuning should not predominantly concentrate on the desired features a fundamental theory is expected to have, but rather on the 
question what a theory needs to qualify as fundamental in the first place. By arguing that a fundamental description of the Universe should possess
zero entropy, we develop a 'holistic' concept for the most fundamental layer of reality: The fundamental description of the Universe is the Universe itself,
understood as an entangled quantum state. 
Adopting a universal applicability of quantum mechanics, in this framework the behavior of subsystems can be understood as the perspectival experience of an entangled quantum Universe perceived through the "lens of decoherence". In this picture the fundamental reality is non-local, and finetuned coincidences in effective theories may be understood in a way similar to EPR-correlations.
This notion provides a fresh view on the topic of Naturalness and Finetuning since it suggests that Finetuning problems and hints for anthropic explanations are an artifact of theories building up on subsystems rather than on the fundamental description.
Recent work in quantum gravity aiming at an understanding of spacetime geometry from entanglement entropy could be interpreted as a  first sign of such a 
paradigm shift. 
\end{abstract}

\section{Introduction}
\label{intro}

Considering its range of validity, its mathematical rigor and the sheer number and immense accuracy of its experimentally verified
predictions, the Standard Model of particle physics easily qualifies as the most successful theory in the entirety of science.
At the same time, however, the Standard Model is also plagued with various Finetuning problems, understood in the remainder of the paper 
as follows:
\smallskip

{\it Finetuning describes unlikely coincidences on an effective level of description which are expected to be explained  
by a more fundamental theory.}

\smallskip
Such unlikely coincidences in the Standard Model include:

\begin{enumerate}

\item
The gauge hierarchy problem: The question why the Higgs mass is so much lighter than the cutoff scale of the Standard Model, arguably 
the Planck scale, corresponding to the unlikely cancellation of quantum corrections to the Higgs mass at the level of $1:10^{17}$
\cite{Giudice:2008bi}.
\item 
The cosmological constant or dark energy problem: The question why the vacuum energy of the Universe is so tiny, i.e. 56-120 orders of magnitude smaller than what would be expected by simple quantum field theory estimates \cite{Martin:2012bt}. 
\item
The strong CP problem: The question why the CP violating $\theta$-term arising naturally in Quantum Chromodynamics is so small
\cite{Dine:2000cj}. 
\item 
The question why the relic density of dark matter is in the same ball park as the relic density of baryons \cite{Wilczek:2004cr}.

\end{enumerate}

In the history of science, such coincidences have always sparked curiosity, and inspired the search for explanations. In this spirit, an overwhelming majority of particle physicists had aspired to new physics at the TeV scale such as supersymmetry or large extra dimensions that could potentially ameliorate 
or solve the gauge hierarchy problem and provide a candidate for dark matter to resolve problem 4 - a strategy 
perceived as 
becoming increasingly unattractive
after the LHC has been running and probing the TeV scale for a decade without finding any physics beyond the Standard Model.
Opinions about what these results implicate for the state and future of fundamental physics differ widely. For one, the present situation has sparked a 
debate
whether "extra-empirical theory assessments" such as simplicity, falsifiability, Naturalness, calculability, or elegance
are reliable guiding principles in the quest for a fundamental theory.  
It is argued whether unnatural theories can be considered to be less attractive as long as no fundamental knowledge about underlying
probability distributions exists - with more \cite{Wells:2018sus} or less optimistic \cite{Hossenfelder:2018ikr} conclusions about the relevance of Finetuning and Naturalness. 

As an alternative, it has become increasingly popular to resort to anthropic explanations justified by various notions of a multiverse of physical realities. Proponents of this idea often argue that Finetuning is no problem in the first place since String Theory predicts a 
multiverse allowing for any set of physical parameters to be realized somewhere among the $10^{500}$ or more vacua in the String Theory ''landscape'' \cite{Schellekens:2013bpa,Bousso:2000xa,Susskind:2003kw}. 
The latter assumption invites for anthropic reasoning, i.e. the speculation that the features of our cosmic environment can't be traced back to
a fundamental theory but rather are contingent upon the condition that they allow conscious observers to evolve.
Such reasoning is, however, extremely controversial:
It has been dismissed, for example, with characterizations such as ''multiverse mania'' \cite{Peter-Woit}, ''Multiverse madness'' \cite{Sabine-Hossenfelder}, ''entirely metaphysical in nature'' \cite{Jim-Baggott},
and the ''most dangerous idea in physics'' \cite{Ellis-Multiverse}, to quote a few.  

On one hand this sentiment is obviously plausible, since anthropic reasoning can prevent a true understanding of phenomena in terms of an underlying theory. On the other hand the existence of a multiverse seems almost inevitable. After all, even if 
String Theory (and the more so the landscape) is at the present stage an extremely ambitious, rudimentary and hypothetical scenario for a theory of everything, the multiverse emerges already in standard quantum mechanics, the paradigm underlying not only the Standard Model of particle physics but almost almost any branch of modern physics: Understanding quantum mechanics in the most straightforward and conservative way as a universally valid theory about Nature, it predicts the parallel realities corresponding to various Everett branches known the "Many Worlds Interpretation"
\footnote{
More concretely: To avoid the conclusion that different Everett branches exist one would have to either alter quantum mechanics for example by adding a process giving rise to a physical collapse, which typically is in conflict with special relativity, or by interpreting quantum mechanics as a theory about knowledge rather than a theory about Nature, assigning it to the realm of humanities rather than science. 
}.
In fact, it has been argued recently that the String Theory landscape and the many worlds multiverse could be one and the same thing, known as the "ER=EPR conjecture" 
\cite{Nomura:2011dt,Bousso:2011up,Nomura:2011rb,Susskind:2016jjb}.
What is often overlooked in this context, however, is that the quantum multiverse is an emergent property of a fundamental Universe
\cite{Wallace:2012zla}.
This latter observation has crucial consequences for our understanding of the fundamental reality. As a consequence, we should  accept that 
it is very likely that the multiverse exists, but focus on the physics beyond the multiverse and ponder whether this may be relevant for the
Finetuning problems of the Standard Model.

In this paper we thus will suggest a fourth avenue:  
It is argued  that not so much the desired properties of a fundamental theory are relevant in this endeavor but rather the
question what qualifies a theory as fundamental in the first place. 
In the following we thus will adopt the assumption that the multiverse is inevitable but emergent and discuss what this 
understanding implies for the notion of a fundamental description of Nature. 
We will start with a discussion of "what is fundamental?" (this part is largely 
identical to the author's entry in the FQXi essay contest of the same name \cite{FQXi-Paes}) before we come back to the question of what this notion means
for the quest for a fundamental theory and the interpretation of the Finetuning problem. 
After all, if Finetuning describes unlikely coincidences on an effective level of description which are expected to be explained by a more fundamental theory, the question how this fundamental theory looks like is a crucial 
ingredient in any solution.

\section{Baryogenesis, Neutrino Mass and the Cosmological Abundance of Dark Matter}

In a nutshell, the basic idea advocated in this essay is that what looks like an unlikely coincidence when observing a subsystem may be resolved by
observing the the total system instead. To illustrate this point, we will first discuss an example where this finding applies
and which is totally unrelated to the topic of quantum foundations discussed later on.

The example presented combines several open questions in particle physics such as the generation of neutrino mass and the origin of the cosmic baryon asymmetry instead, and it is related to what in the good old days of the late 1990ies - before the advent of dark energy - was known as the "cosmological finetuning problem" \cite{Chankowski:1998za}, i.e. problem 4.

Neutrino masses are generally understood as the first laboratory hint for particle physics beyond the Standard Model, and the reason is that they are
deeply connected with the question whether lepton number is violated or conserved. Neutrino masses can be realized in two general ways, either
the neutrino is coupled to a new, gauge singlet right-handed neutrino, ${\cal L}_{\rm Dirac} = m_D  \overline{\nu}_L \nu_R + hc$,
or to its own anti-particle,  ${\cal L}_{\rm Majorana} = m_M \overline{\nu}_L \nu^c_R $. While the latter option violates lepton number, giving rise
to various new processes forbidden in the Standard Model, the second option does not. It introduces, however, a new particle $\nu_R$ whose
Majorana mass ${\cal L'}_{\rm Majorana}=m_M \overline{\nu}_R \nu^c_L $ is allowed by the Standard Model gauge symmetry. This mass term would
also violate lepton number, implying that the discovery of neutrino masses either implies that lepton number is broken or that there is some new
symmetry or principle guaranteeing lepton number conservation and thereby forbiding the right-hganded Majorana mass ${\cal L'}_{\rm Majorana}$. The most popular neutrino mass models assume that lepton number indeed is violated since this allows for elegant explanations why the
neutrino mass is so small, such as the various versions of the seesaw mechanism \cite{Minkowski:1977sc}. 
Lepton number violation can be cosmologically dangerous, though. 
Understood as breaking of the anomaly-free combination $B-L$, it can erase any pre-existing baryon asymmetry when being combined with
the non-perturbative $B+L$ violating sphaleron processes present within the Standard Model \cite{Deppisch:2015yqa}. 
This reasoning seems to exclude
lepton number violation in the range where sphaleron processes are active (unless the lepton number violation is directly related to the
origin of the baryon asymmetry such as in vanilla-type leptogenesis models \cite{Fukugita:1986hr}), 
thereby ruling out many attractive neutrino mass models.
The conclusions holds, however, only as long as one is concentrating only on leptons and baryons. In asymmetric dark matter models the
production of baryons and dark matter is related - offering a possibility to explain Finetuning problem 4. With respect to the inconsistency of 
successful baryogenesis and lepton number violation,
such models supply the early Universe with a "third reservoir" (leptons and baryons plus dark matter) to the cosmic soup 
and thereby invalidate the argument introduced above
\cite{Frandsen:2018jfi}. This simple example demonstrates that what looks
like finetuned and inconsistent when concentrating on a subsystem (leptons and baryons), may turn out to be perfectly natural when looking at the total system (leptons, baryons and dark matter). 
While a pencil balancing on its tip seems unnatural, the pencil's state can be explained naturally when including the writer's hand holding it.
Although the subsystem (pencil) seems to behave unnatural, the total system (pencil + hand) is not. 

But how is the total system related to the fundamental theory supposed to explain the unlikely coincidences understood as Finetuning problems
on the effective level of description? In the following we will argue that the most fundamental level of description {\it is} indeed represented by the state of the total system rather than by the states of constituents. To come to this conclusion, we will develop  a consistent notion of a fundamental theory. In this context we can argue consistently
that quantum mechanics suggests that this fundamental theory can only be the Universe itself. Finally, we will come back to the problem of Naturalness from this perspective.


\section{What is Fundamental?}

{\it "There is only one world, the natural world... There are many ways of talking about the world... 
Our purposes in the moment determine the best way of talking."}
(Sean Carroll, \cite{Carroll})
\medskip

We don't live in a single reality - we live in many realities. And maybe exactly this feature is what makes us human. As the Israeli historian Yuval Noah Harari
argues:

{\it  "...the truly unique feature of our language is not its ability to transmit information about men and lions. Rather, it's the ability to transmit information about things that do not exist at all... It's relatively easy to agree that only Homo sapiens can speak about things that don't really exist, and believe six impossible things before breakfast... law, money, gods, nations....
Any large-scale human cooperation -- whether a modern state, a medieval church, an ancient city or an archaic tribe -- is rooted in common myths that exist only in people's collective imagination. The kinds of things that people create through this network of stories are known in academic circles as 'fictions', 'social constructs' or 'imagined realities'...
The ability to create an imagined reality out of words enabled large numbers of strangers to cooperate effectively...
This opened a fast lane of cultural evolution, bypassing the traffic jams of genetic evolution"} \cite{Harari}.

Without any doubt this hypothesis is intriguing. But can we really deny entities such as money, stock corporations or law to be part of reality? After all, a lack of money, a crash at the stock market or a violation of law can get us in serious trouble. Moreover, if Harari argues such social constructs exist only in people's imagination, what about people themselves? What about biological organisms?

When Erwin Schr\"odinger tried to define life, he described it as a property \cite{WhatisLife}:

{\it "What is the characteristic feature of life? When is a piece of matter said to be alive? When it goes on 'doing something', moving, exchanging material with its environment, and so forth, and that for a much longer period than we would expect an inanimate piece of matter to 'keep going' under similar circumstances."}

Schr\"odinger's characterization of life by fiat of its striking stability supported the hypothesis of an hereditary molecule which paved the way for the field of molecular biology and later on motivated James Watson and Francis Crick to decode the structure of DNA \cite{DNA}. Moreover, it defines life by its functionality rather than by its material basis, as an information processing routine rather than an object in space and time. This view is supported by the fact that over a typical human lifespan most atoms in the human body will be replaced without altering the identity of the individual.
In this sense biological organisms resemble much more things we call social constructs than matter itself.
So after all, if we deny corporations to be part of reality we have to deny reality to ourselves as well.

Once we embrace the concept of many realities, on the other hand, new questions arise. 
Philosophers talk about weak or strong emergence when the natural laws of one level of description (such as biology) can not - either in practice or fundamentally - be derived from laws on a more fundamental level such as physics. While the existence of weak emergence is unchallenged - nobody wants to describe the stock market by calculating the behavior of the elementary particles involved - accepting strong emergence on the other hand is not
quite different from believing in miracles. After all, if phenomena on a higher, more complex level are not even in principle describable on a more fundamental level, then the fundamental level does also not constrain the space of
possibilities of the higher level. For example, if the fact that particles can't propagate faster than light does not imply that also biological organisms can't travel faster than light, there is also nothing which prevents Jesus from
walking on water. Any extrapolation of known physical laws, meaning any application of these laws in new situations, would become questionable. In this case the entire scientific endeavor wouldn't make sense any more.

Once we exclude the possibility of strong emergence it is clear that not all realities are equally fundamental.
Rather the more fundamental realities constrain the space of possibilities for the higher levels: While more fundamental realities or natural laws are still valid on higher levels, more complex or higher descriptions have a limited range of applicability when being extrapolated to more fundamental constituents. This view suggests a hierarchy of sciences similar to the one postulated by the french philosopher Auguste Comte. In this hierarchy physics defines the foundation, chemistry is the physics of the outer atomic orbits, biology deals with the chemistry
of complex organic molecules, psychology describes the biology of the neural system and sociology and economics discuss the psychology of large numbers of individuals.

While it has been pointed out recently by Erik P. Hoel and collaborators \cite{Hoel1,Hoel2}, that higher levels of description may actually be more deterministic (a typical example is that macroscopic classical physics appears to be more deterministic than microscopic quantum mechanics), it seems clear that realities are the more fundamental, the more observer-independent they are. Conversely, higher layers of description are typically build on concepts
which are more substrate independent, relying more on the processing of information than on the properties
of their actual constituents and in this sense are also more idealized. Obviously the science of sociology makes
only sense from a viewpoint which accepts the existence of biological organisms beforehand, while particle physics in contrast
does not rely on any such premise. 

Already at this point however we realize that we have to revise our traditional concept of reductionism. Once we reduce sociology and psychology to biology it is obvious that
reductionism can no longer be understood as a materialistic approach, but rather a concept in information theory: Rather than identifying constituents, reductionism reduces complex systems by identifying the most relevant degrees of freedom on the higher layer of reality with respect to the degrees of freedom on a more fundamental level.

\section{Zeh versus Democritus}

Naively the above mentioned scheme seems  to suggest an ontology based on particles, a traditional reductionist view in which ever smaller constituents 
determine the properties of the higher, more complex layers.  An ontology resembling the lore of the
the pre-socratic Greek philosopher Democritus:
{\it "Only the atoms and the void are real"}\cite{Democritus}.
Such an ontology would indeed be appropriate for a classical world. In quantum mechanics, however, 
this notion fails, as has been 
emphasized by H. Dieter Zeh \cite{Zeh:1992qd}:
{\it "There are no quantum jumps, nor are there particles!"}. As Zeh points out: {\it "Quantum theory does not require the existence of discontinuities: neither in time (quantum jumps), nor in space (particles), nor in spacetime (quantum events). These apparent discontinuities are readily described objectively by the continuous process of decoherence."}

The process of decoherence, discovered by Zeh in 1970 \cite{Zeh:1970zz}, describes a phenomenon which is
well understood theoretically and experimentally confirmed \cite{Haroche}: Whenever a quantum system is measured or coupled to its environment, the system gets entangled with both observer and environment. As the environment is not totally known to the observer, this leads to the loss of information about the quantum system getting delocalized into the environment. As a consequence, quantum superpositions -- such as Schr\"odinger's infamous undead cat \cite{Cat} -- decay rapidly and quasi-classical objects such as particles with a definite location emerge. An intuitive metaphor for the process of decoherence is the 
action of a
colored optical lens. While it seems that such a lens adds color to the colorless sunlight, in truth the 
lens works by absorbing all other component colors being present in the white, colorless state. Just like decoherence, 
a colored lens seemingly creates information by actually filtering out information.

Note, that quantum mechanics is understood here as a paradigm, not as a reference to first-quantized particle quantum mechanics. In this sense quantum field theory (which indeed can be understood as the physics of an array of coupled quantum-mechanical harmonic oscillators is an example of quantum mechanics at work. Any deeper understanding of the meaning of quantum mechanics will also apply to quantum field theory.
In rapport with Zeh's understanding, particles are usually understood as field excitations in quantum field theory.
Moreover, as exposed by the Unruh effect \cite{Unruh}, particles are not observer-independent entities: According to quantum field theory an accelerated observer should observe a blackbody radiation in vacuo. 
(More accurately, particle number is not observer-independent. Attributes of the quantum field such as rest mass, spin, and other quantum numbers of course are.) 

Adopting Occam's razor, one can assume now, that the process of decoherence is entirely responsible for the quantum-to-classical transition which leads to various decohered quasi-classical realities, the so-called Everett branches \cite{Everett:1957hd}. This interpretation of 
the quantum measurement process is usually known as the Many-Worlds-Interpretation, Many-Minds-Interpretation or Universal Quantum Mechanics.

\section{A crisis of reductionism?}

These observations expose two interesting facts which are usually overlooked in the traditional notion of reductionism: 

First, according to quantum theory, at some point the naive notion of reductionism, in the sense 
that emergent properties of higher layers of description can be at least in principle explained by the 
properties of isolated constituents, breaks down, as the mere existence of these constituents itself turns out
to be a consequence of isolating the corresponding degrees of freedom from their environment.
In other words, the existence of particles is a consequence of a specific perspective:
Whenever a particle is observed, the quantum field gets coupled to the measurement device and its unknown environment. It is this interaction and the information loss that implies that the state of the quantum field is observed as a particle. Particles thus are a consequence of decoherence. Particles are, as Zeh argues
consistently, emergent themselves (in the sense that a description in terms of constituents neglects information about the entangled total system). 

And second, while more fundamental concepts or realities seem on one hand to be more observer-independent,
they are on the other hand often not directly observable anymore. In quantum mechanics, the quantum mechanical wave function is unanimous for any observer, while a definite classical result of a measurement applies to one
specific Everett branch, only. Another example is the fundamental role symmetries play for example in
particle physics. As has been argued by Werner Heisenberg in 1932 \cite{Isospin}, the protons and neutrons constituting the atomic nucleus behave so similar that
they can be understood as two states of a single particle -- in analogy to a quantum spin which can point 
up and down. Heisenberg thus described proton and neutron together as the two possible states of an isospin doublet. Curiously, this way the 
apparent material difference of proton and neutron is reduced to the information about the state of the
isospin Q-Bit. 

Over the last decades, particle physicists mostly followed a paradigm suggesting that a more fundamental description corresponds to more symmetry.
The concept of a Grand Unified Theory, for example, generalized Heisenbergs isospin symmetry, described by the Lie group SU(2) to enlarged groups such as SU(5) \cite{SU5} or SO(10) \cite{SO10} where all fundamental matter constituents are understood as different states of one or two basic representations ("basic" does not necessarily refer to the fundamental representation, here. In SO(10) GUTs the basic representation is the 16-dimensional spinor representation, in SU(5), it refers to the fundamental and the antisymmetric 10-dimensional representations.).

Now, do these observations imply that the paradigm of reductionism is in trouble? It isn't, once we 
decide to understand reductionism as an information theoretic approach.
As has been argued above, on the one hand this decision shifts many phenomena usually understood as materialistic -- 
such as the properties of elementary particles -- into the immaterial realm of information.
On the other hand, so far there is no evidence that information can exist without a material medium
or carrier. Quite contrary, whenever we encounter information this information seems to depend on
how we interpret or view a materialistic object: We simply cannot extract a Beethoven symphony from
 a USB stick without a device which can read out the stick and process the data format appropriately.
 
 So while the relations between information and matter are subtle, our perspective onto the world seems to play a crucial role, here.

\section{How to quantify fundamentality?}

Let us come back now to the original question: What is fundamental?
As we have seen, fundamental descriptions in modern physics sacrifice the requirement of direct observability
for the sake of an observer-independent description. But how can observer-independence of a theory be quantified?

To discuss this point, we turn to the prototypical example of an emergent theory: thermodynamics in its incarnation as statistical mechanics. In thermodynamics, states such as gases or liquids are described by parameters or state functions such as temperature, pressure or volume. These state functions  are not fundamental in the sense that they do not correspond to a specific configuration of the constituent atoms or molecules (a so-called "microstate"), but to a statistical average of microstates known as "macrostate". For example, the temperature of a macrostate
can be related to the average energy of the constituent microstates. The normalized logarithm of the number of microstates corresponding to a given macrostate is known as entropy. Thus entropy can be understood as the missing information to identify the exact microstate in a given macrostate and determines how probable 
the macrostate is. According to the second law of thermodynamics, entropy also determines the direction of
thermodynamic out-of-equilibrium processes (towards higher entropy) and thus establishes an (or presumable the) arrow of time 
of the Universe.

The concept of entropy also allows us to find a concept of fundamentality which is more general than the 
spoiled constituent concept of traditional reductionism. A microstate is fundamental, a macrostate is not.
Consequently, the fundamental state of the Universe has zero-entropy (and is arguably timeless).
In fact, the Wheeler-DeWitt equation  \cite{DeWitt:1967yk} of canonical quantum gravity describes a timeless
Universe on the fundamental level \cite{Zeh:1986ix,Kiefer:2009tq}.

Turning back to quantum mechanics, it is well known that in the quantum-to-classical transition the von-Neumann entropy increases as a consequence of the information loss into the environment. Thus the fundamental state
of the Universe can not be a constituent, it has to be the total entangled system of observer, measured system,
and environment, also known as the quantum Universe itself. 
Similar conclusions have been drawn previously from a philosopher's perspective by Jonathan Schaffer \cite{Schaffer} and elaborated recently by Claudio Calosi \cite{Calosi} under the term "quantum monism".

\section{Lessons for the search for a  Theory of Everything}

The realization that no local constituent can really be fundamental leads to several interesting consequences
for our search for a Theory of Everything (TOE). As such a TOE should identify the fundamental reality of the Universe, it should be questioned whether the traditional reductionist approach -- which so far beyond any
question was amazingly successful -- will be still promising once we approach the fundamental level of reality.
Will the hierarchy of sciences starting from  Sociology and Economy, over Psychology, Biology, Chemistry
to Atomic physics, Nuclear Physics and Particle Physics will continue towards String Theory, where particles are described as excitations of localized, one-dimensional objects? 
Or will it be more promising to try to derive the Standard Model of particle physics directly from a non-local 
concept such as the wave function of the Universe in Quantum Cosmology? 
Indeed, in my opinion strings do not qualify as a fundamental description of the Universe. String Theory, however, has also given rise to a variety of non-local concepts which could indeed be instrumental in establishing a description in terms of a unique fundamental quantum state, such as gauge-gravity duality, holographic screens and gravity emerging from entanglement.
Could this be obtained by
identifying our perspective onto the Universe, by realizing that:
\\ 
\\
{\bf Reality = Universe + Perspective}\\
\\
and that
\\
\\
{\bf Fundamental Reality = Universe without Perspective}?\\
\\
Can the notion of information being processed in higher layers of reality be traced back to this perspective?
\footnote{textcolor{red}{What is meant here is that observer-independence corresponds to a description that does not rely on a specific perspective. This does not imply that there are no observers.}} 

Or, phrased as ideas to inspire an experimental program:

Can the entanglement of the Quantum Universe be verified by finding unexpected correlations, similar to 
the ones found in EPR \cite{EPR} like experiments,
for example in the Cosmic Microwave Background \cite{CMB}? Is our perspective onto the Universe related to the way our consciousness 
works \cite{Paes}, for example if we can show that consciousness is a by-product of a certain algorithm running on the
localized degrees of freedom in our brains? How for example would the Universe look like for example for an alien intelligence whose
consciousness may be a product of an algorithm running on momentum eigenstates such as plane waves?
Quite obviously, in this case decoherence may converge towards a different "preferred basis" \cite{Zurek}.
And finally, are symmetries as a concept based on invariance related to us observing a macrostate and thus only
seemingly observer-independent \cite{Symmetries,Witten:2017hdv}? Or is the fundamental reality totally symmetric and thus essentially featureless 
\cite{Tegmark} and
everything we observe in the Universe turns out to be a consequence of information which can eventually be traced back to
our perspective onto the Universe?

\section{A fundamental Universe}

To summarize,
the question "what is fundamental?" leads to interesting insights about the notion of reality.
As we argued above, a consistent concept of reductionism has to interpret the reduction of higher 
levels of reality such as societies or life into lower levels such as atoms or quantum fields as an
information theoretic approach where more fundamental layers of reality are characterized by lower entropy. On the one hand information itself is immaterial - people have been killed by rocks but nobody ever has been killed by a Beethoven symphony (unless for example it was scribbled on a rock).
On the other hand, information relies - as far as we know - on a material carrier and depends on the perspective of the observer. The question which
properties of the Universe are related to information and which ones are related to matter may well be one of the most pressing problems 
dominating physics and information science of the 21st century. 
If some fundamental carrier of information exists, it probably will be possible to be traced back to the fundamental level of reality, itself, 
the zero-entropy microstate which according to quantum physics - at least as far as quantum mechanics is 
universally applicable -- is the quantum state of the Universe itself. Such a state would include not only
the visible Universe, dark matter and dark energy and other possible parallel Universes which could be
produced in primordial inflation \cite{Inflation}, it also includes other possible realities corresponding to different Everett
branches and all their superpositions. Everything above would dissolve in a single entity described by an 
entangled quantum state which only from a certain, local perspective would look like a network of interacting quantum fields and particles. If the statement that  the fundamental layer of reality is the Universe itself sounds tautological, that is,
since the 'Universe' usually is understood as a multiplicity of things, including particles, fields, stars, planets and lifeforms. 
On the contrary,
the fundamental Universe is a single entity which only looks like many things as perceived through the lens of decoherence.
Physics is approaching Metaphysics here, and who wants can find interesting parallels to the concept of such an all-encompassing reality for example in the philosophy of Neoplatonism, east asian philosophies or religion.

\section{Fine Tuning and Perspective in an Entangled Quantum Universe}  

We now are finally coming back to the 
Finetuning problem of the Standard Model and the issue of Naturalness as a guiding principe. So far we have argued that in order
to discuss whether unlikely coincidences on an effective level of description are resolved in a more fundamental theory, one first has to specify what
"fundamental" really means. This discussion led to the conclusion that the fundamental description of the Universe has to be a single, entangled quantum state. But is this observation really helpful to deal with the Finetuning problems encountered in particle physics and cosmology?
This is admittedly hard to say. There exist, however, a few hints that indeed point towards the existence of such a relationship which will be summarized in the following.

The main point advocated in this essay is that the Finetuning problems in our best theories are not so much deficits of a or a group of specific 
theories bur rather a symptom of the concept of "fundamentality" underlying these theories. In the following we will try to make this statement
somewhat more precise: We will provide some arguments for the conclusion that the nature of Naturalness problems may hint at an insufficient understanding of quantum mechanics,
that the likely existence of a multiverse does not necessarily suggest anthropic reasoning, and that first concrete examples of the suggested
paradigm shift can be observed already in recent approaches to quantum gravity.

\begin{itemize}

\item
First, we observe that
the notion of Naturalness is strongly related to the concept of effective field theories: In this context the possibility to integrate out short-distance degrees of freedom allows to describe large scale physics without knowing the exact UV completion. This implies that Naturalness is strongly
related to the question whether physics at large and small scales can be separated and described independently. Conversely, quantum physics and in particular entanglement are inherently non-local: For entangled systems, the small subsystems are emergent upon the total large system.
The fact that an absence of Naturalness is on one hand related to a genuine quantum phenomenon such as renormalization and that it spoils the separation of scales by integrating out short-distance degrees of freedom in effective field theories while on the other hand quantum physics has inherent non-local properties could point towards an insufficient understanding of quantum mechanics as the underlying problem. After all, quantum mechanics has been hurriedly applied to many phenomena in solid states and atomic, nuclear and particle physics without finding an interpretation generally agreed on.

\item
Next,
while universal quantum mechanics predicts a multiverse of alternative realities or Everett branches, that doesn't mean that it necessarily suggests anthropic explanations for fundamental laws of Nature. Of course a dolphin may argue righteously that it lives in the sea since it couldn't survive for long outside of water. But there exists a more profound explanation why the dolphin encounters water in its environment which 
relies on the origin of elements in Big Bang and Supernova nucleosynthesis, chemistry, structure formation, Earth's gravitational pull and its
distance to the sun. Similarly, quantum mechanics can give rise to seemingly unlikely correlations which nevertheless are explained 
by a global property of an entangled state. 

Consider, for example, the EPR correlations \cite{EPR} of the Bell singlet state, 
\begin{eqnarray}{}
| \psi_{s} \rangle = | \uparrow \downarrow \rangle -  | \downarrow \uparrow \rangle.
\end{eqnarray}
When measuring the polarization of one of the constituent spins every result is equally likely. For example, when measuring the
first spin along the z-axis two Everett branches emerge, one with spin up and one with spin down, and the observer will find herself with
an equal probability of 50\% in either of these branches. Both alternatives are realized in the Everettian quantum multiverse. However,
when comparing measurements on both constituent spins, while their individual directions remain undetermined, they will always add up to zero in any Everett branch.
Thus it is obvious that even in multiverse scenarios there typically exists "physics beyond the multiverse", meaning global properties that
are realized in any of the parallel universes - at least in the context of the Many Worlds Interpretation. Moreover, as has been mentioned above,
recently Nomura, Susskind, Bousso and others have argued that the quantum multiverse and the multiverse arising in chaotic inflation 
(realizing the String Theory landscape) could turn out to be one and the same thing \cite{Nomura:2011rb,Bousso:2011up,Nomura:2011rb,Susskind:2016jjb}. 
In this case the argument favoring physics beyond the multiverse applies equally well for other
bubble universes in cosmology and competing String Theory vacua.

\item
Finally, there exists a first interesting, recent approach which seems to derive successfully seemingly fundamental aspects of Nature as
emergent properties arising by observing a subsystem of an entangled quantum state:
As Mark van Raamsdonk has recently pointed out, spacetime geometry can be derived from the entanglement entropy of a quantum field theory 
defined on the boundary of the spacetime considered \cite{VanRaamsdonk:2010pw}.
In this context, van Raamsdonk’s work provides an interesting and promising example pointing out concretely how physics could emerge from entanglement. It is thus a concrete example how a theory building on the ideas developed in this paper could look like. 
While many open questions remain, it is conceivable that a similar mechanism
could give rise, for example, to gauge symmetries. In this case van Raamsdonk's and other's related work could be the first harbinger
of a paradigm shift related to our notion of what should be understood as a fundamental theory.

\end{itemize}

\section{Summary}

This paper tries to provoke a fresh view onto the problem whether - and if so how - Naturalness and Finetuning can guide us in our search for a fundamental description of Nature. It argues that this endeavor should not so much concentrate on desired features of a fundamental theory but rather on what qualifies a theory as fundamental in the first place. In this context zero-entropy is argued for as the most relevant criterion for fundamentality and it is shown that - adopting a universal validity of quantum mechanics - only the Universe itself, understood as a single, entangled quantum state, can be considered as fundamental. The suggestion that the total rather than the constituent system represents the fundamental description of Nature
may turn out to be highly relevant for the discussion of Naturalness and Finetuning, since there exist many examples where what looks finetuned or  
inconsistent when analyzing a subsytem turns out to be totally natural when looking at the total system instead. 
It thus should be critically scrutinized
whether
the quest for a fundamental theory should commence from the viewpoint of quantum cosmology rather than with particles or strings, and whether particle physics could be derived from entanglement and the observer's perspective
onto the total, fundamental quantum state of the Universe. Recent success in describing spacetime geometry as a consequence of entanglement entropy \cite{VanRaamsdonk:2010pw}
suggest that such an approach
may indeed be promising. As
Gian Giudice has argued \cite{Giudice:2017pzm}: 
{\it ''we live in times of great uncertainties – the
best moments for scientific revolutions to happen...
we are confronted with the need
to reconsider the guiding principles that have been used for decades to address the most
fundamental questions about the physical world''}. 
It is quite possible that the dawn of such a paradigm shift is already breaking.

\end{document}